\documentstyle[psfig,conf_iap,]{article}

\newcommand{\ovi}{O~{\sc vi}\relax}
\newcommand{\civ}{C~{\sc iv}\relax}
\newcommand{\feii}{Fe~{\sc ii}\relax}
\newcommand{\halpha}{H$\alpha$}

\newcommand{\msun}{M$_\odot$}
\newcommand{\kms}{km~s$^{-1}$\relax}

\newcommand{\column}{cm$^{-2}$}

\newcommand{\fuse}{{\em FUSE}}
\newcommand{\rosat}{{\em ROSAT}}

\begin{document}
\heading{Interstellar \ovi\ in the Large Magellanic Cloud} 

\par\medskip\noindent

\author{J. Christopher Howk$^{1}$}
\address{Department of Physics and Astronomy,
The Johns Hopkins University, Baltimore, MD, 21218; howk@pha.jhu.edu }

\begin{abstract}

I summarize {\em Far Ultraviolet Spectroscopic Explorer} (\fuse)
observations of interstellar O~{\sc vi} absorption towards 12
early-type stars in the Large Magellanic Cloud (LMC), the closest disk
galaxy to the Milky Way.  LMC \ovi\ is seen towards all 12 stars with
properties (average column densities, kinematics) very similar to
those of the Milky Way halo, even though O/H in the LMC is lower by a
factor of $\sim2.5$.  Sight lines projected onto known LMC
superbubbles show little enhancement in \ovi\ column density compared
to sight lines towards quiescent regions of the LMC.  The
\ovi\ absorption is displaced by $\sim-30$ \kms\ from the corresponding 
low-ionization absorption associated with the bulk of the LMC gas.
The LMC \ovi\ most likely arises in a vertically-extended
distribution, and I discuss the measurements in the context of a halo
composed of radiatively-cooling hot gas.  In this case, the mass-flow
rate from one side of the LMC disk is of the order $\dot{M}
\sim 1$ \msun\ yr$^{-1}$.

\end{abstract}

\section{Introduction}

The production of hot, highly-ionized gas in galactic environments is
closely related to the input of energy and matter from stars and
supernovae into the interstellar medium (ISM).  Such ``feedback'' can
shape the ISM on kiloparsec scales in regions with high concentrations
of early-type stars.  In disk galaxies, with differing pressure
gradients in the vertical and radial directions, such energy input is
responsible for the production of vertically-extended ``halos'' or
``coronae'' about these systems \cite{deavillez00,normanikeuchi89}.

Howk et al. \cite{howk02} have recently completed a study of
interstellar \ovi\ in the Large Magellanic Cloud (LMC).  Because the
ionization energy required for its creation (IP$_{\rm O\, V - O\, VI}
= 114$ eV) precludes its production via photoionization by starlight,
\ovi\ is a tracer of hot ($\sim3\times10^5$ K), collisionally-ionized
gas in galactic environments.  Therefore, the Howk et al. study of
\ovi\ in the LMC provides fundamental information on the content,
distribution, and kinematics of material created by the interactions
of stars and supernovae with the ISM in the closest and best-studied
disk galaxy beyond the Milky Way.  I summarize the principle results
of this study below.

\section{LMC \ovi\ -- Content and Kinematics}

\begin{figure}[t]
\centerline{
\vbox{\psfig{figure=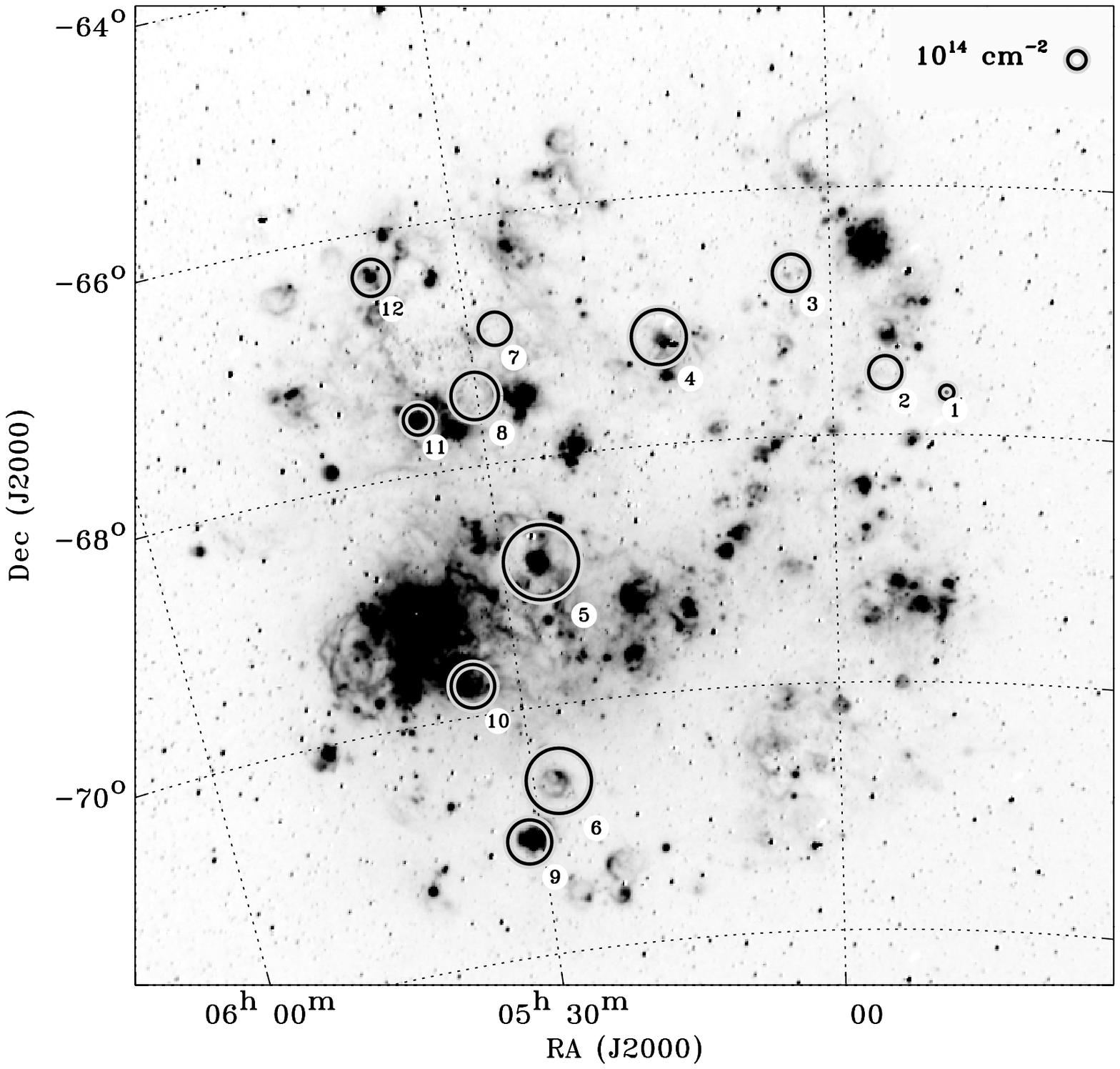,width=7.cm}}
\vbox{\psfig{figure=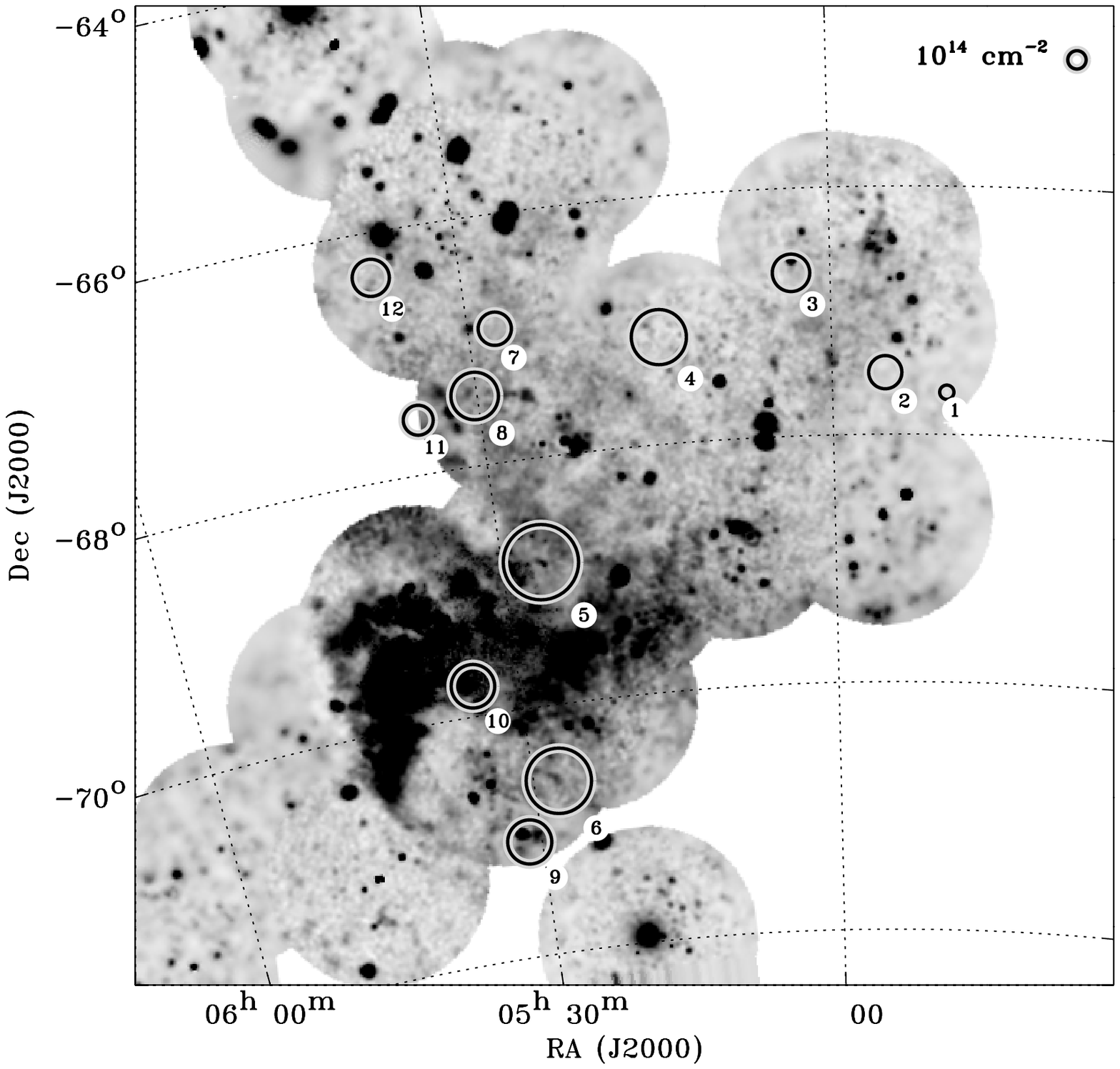,width=7.cm}
}}
\caption[]{\halpha\ image ({\em left}) of the LMC \cite{gaustad01}
and 0.5-2.0 keV \rosat\ PSPC mosaic ({\em right}) of the LMC 
\cite{snowdenpetre94}  with positions of the \fuse\ probe stars 
marked.  The radius of each circle is linearly proportional to the
column density of \protect\ovi\ at LMC velocities (scale given in the
upper right).  
\label{fig:maps}}
\end{figure}

Table 1 summarizes the statistical properties of the
\ovi\ column densities, $N(\mbox{\ovi})$, along the  12 sight lines 
in the Howk et al. \cite{howk02} study, while Figure \ref{fig:maps}
compares the distribution of LMC \ovi\ with \halpha\
\cite{gaustad01} and \rosat\ hard X-ray \cite{snowdenpetre94} views of
the LMC.  Significant \ovi\ is observed across the whole face of the
LMC, although the \ovi\ is patchy: the standard deviation of the
measurements is $38\%$ of the mean.  Sight lines projected onto known
superbubbles (the four sight lines south of $\delta = -68^\circ$ in
Figure \ref{fig:maps}) show only very modest (if any) \ovi\
enhancements compared with more quiescent-appearing sight lines (e.g.,
Nos. 1--4 in Figure
\ref{fig:maps}).


\begin{center}
\begin{tabular}{ll}
\multicolumn{2}{c}{{\bf Table 1.} Statistical Properties of}  \\
\multicolumn{2}{c}{Interstellar \ovi\ in the LMC} \\
\hline
\multicolumn{1}{c}{Quantity} &
\multicolumn{1}{c}{Value} \\ 
\hline
$\langle N(\mbox{\ovi}) \rangle$ & $2.34\times10^{14}$  \column \\
$\sigma [ N(\mbox{\ovi})] $      & $0.89\times10^{14}$  \column \\
$\langle N_\perp (\mbox{\ovi}) \rangle$ 
			         & $1.95\times10^{14}$  \column \\
\hline
\end{tabular}
\end{center}


For comparison with the Milky Way values reported by Savage et
al. \cite{savage00}, we define the quantity $N_\perp (\mbox{\ovi})
\equiv N(\mbox{\ovi}) \, \cos i$, where $i$ is the inclination of the
LMC.  This is the column density projected perpendicular to the plane
of the LMC and is equivalent to the $N_\perp \equiv N \sin |b|$
measurements in the Galaxy.  The first \fuse\ measurements of Galactic
halo \ovi\ towards extragalactic objects found $\langle N(\mbox{\ovi})
\sin |b| \rangle = 14.29$ ($38\%$ standard deviation).  The average LMC and
Milky Way $N_\perp (\mbox{\ovi})$ values -- and their standard
deviations -- are identical.


\begin{figure}[t]
\centerline{
\vbox{\psfig{figure=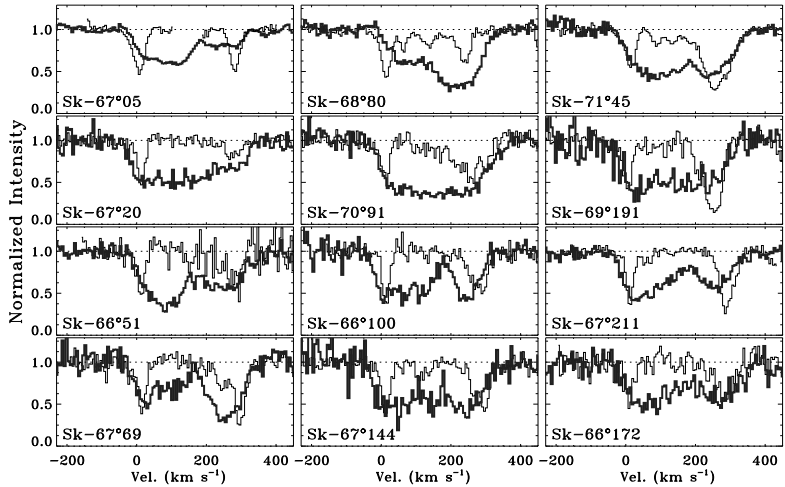,width=4.0in}}}
\caption[]{The observed absorption line profiles of \ovi\
$\lambda 1031.926$ (thick line) and \feii\ $\lambda 1125.448$ (thin
line) towards the LMC stars studied in Howk et al. \cite{howk02}.  The
\feii\ absorption components associated with the Milky 
Way ($v\lsim +175$ \kms) and LMC ($v \gsim +175$ \kms) are seen to be
significantly narrower than the corresponding \ovi\
profiles.\label{fig:profiles}}
\end{figure}

The observed kinematic profiles of the LMC \ovi\ are quite broad, with
breadths implying $T \lsim (2-5)\times10^6$ K.  Figure
\ref{fig:profiles} shows the normalized \ovi\ absorption profiles along
the 12 sight lines studied and the profiles of a moderate-strength
\feii\ transition.  The latter traces gas associated with  neutral 
material in the LMC disk.  The \ovi\ is much broader than the
\feii\ along all sight lines and is systematically blue-shifted from 
the disk (by $\sim30$ \kms\ on average).  Thus, the \ovi\ is
kinematically decoupled from the bulk of the LMC disk material.  There
is gas present at velocities compatible with the outflow of material
from the LMC disk along all of the sight lines discussed by Howk et
al.  In only two cases is there possibly \ovi\ at velocities that may
indicate infall.

\section{Interpretation}

The \fuse\ observations reveal interstellar \ovi\ associated with the
LMC is present in large quantities across the whole face of the LMC,
with an average column density and patchiness identical to those of
the Galactic halo.  Lines of sight projected onto superbubbles and
supergiant shells have much the same column densities as lines of
sight projected onto quiescent regions.  The LMC \ovi\ absorption is
both much broader and shifted to lower absolute velocities than the
lower-ionization gas (e.g., \feii).

For reasons discussed in detail by Howk et al. \cite{howk02}, the
favored interpretation of these salient aspects of the LMC \ovi\ is
that the LMC is surrounded by a hot, highly-ionized halo or corona --
similar in many respects to that found in the Milky Way -- that gives
rise to the observed \ovi\ absorption.  Several models can explain the
physics of the \ovi\ production within a gaseous halo about the LMC,
including cooling galactic fountain flows and interface models (such
as turbulent mixing layer or conductive interface models).

The cooling fountain model provides an elegant explanation for the
similarity of the average Milky Way and LMC \ovi\ column densities.
Though these galaxies differ in oxygen abundance by a factor of
$\sim2.5$, the column density of highly-ionized metals in a cooling
flow of hot material is independent of abundance
\cite{edgarchevalier86}.  The column density of \ovi\ in the Edgar \&
Chevalier \cite{edgarchevalier86} models is a function of the ratio
$\dot{N}/n_0$, where $\dot{N}$ is the cooling rate (in protons
\column\ s$^{-1}$), and $n_0$ is the initial density of the flow.  The
average LMC \ovi\ column density (Table 1) corresponds to a one-sided
mass-flow rate from the LMC disk of
\begin{displaymath}
\dot{M} \sim 1 \, \left( \frac{n_0}{10^{-2} \ 
	{\rm cm^{-3}}} \right) {\rm \ M_\odot \ yr^{-1}}.
\end{displaymath}
The adopted density is consistent with estimates of electron densities
in supergiant shells and diffuse gas using X-ray observations of the
LMC \cite{points01}.

It should be noted, however, that the energy input requirements into
the ISM and mass flow rates from the disk can be significantly
different if the \ovi\ arises in turbulent mixing layers or other
interface-type models.  Observations of other highly-ionized species
(e.g., \civ) will be required to distinguish between the cooling flow
and interface models.

\acknowledgements{Thanks to my collaborators for all of their help
in this project.  The LMC \halpha\ image is from the Southern H-Alpha
Sky Survey Atlas (SHASSA), which is supported by the National Science
Foundation. This work is based on data obtained for the Guaranteed
Time Team by the NASA-CNES-CSA FUSE mission operated by the Johns
Hopkins University. Financial support to U.S. participants has been
provided by NASA contract NAS5-32985.  I also recognize support from
NASA Long Term Space Astrophysics grant NAG5-3485 through the Johns
Hopkins University.}

\vfill
\end{document}